\documentclass[12pt]{article}
\usepackage{amssymb}
\usepackage{amsmath}
\usepackage{graphicx}

\newcommand{\npb}[3]{Nucl.~Phys. {\bf B#1} (#2) #3}
\newcommand{\plb}[3]{Phys.~Lett. {\bf B#1} (#2) #3}
\newcommand{\prl}[3]{Phys.~Rev. Lett. {\bf #1} (#2) #3}
\newcommand{\prd}[3]{Phys.~Rev. {\bf D#1} (#2) #3}

\newcommand{\cqg}[3]{Class.~Quant.~Grav. {\bf #1} (#2) #3}

\newcommand{\jhep}[3]{JHEP {\bf #1} (#2) #3}

\renewcommand{\hbar}{\overline{h}}

\newcommand{\be}{\begin{equation}}
\newcommand{\bea}{\begin{eqnarray}}

\renewcommand{\d}{\partial}
\newcommand{\ee}{\end{equation}}
\newcommand{\eea}{\end{eqnarray}}

\newcommand{\sect}[1]{\setcounter{equation}{0}\section{#1}}

\newcommand{\al}{\ensuremath{\alpha}}

\newcommand{\vep}{\ensuremath{\varepsilon}}

\newcommand{\La}{\ensuremath{\Lambda}}

\renewcommand{\d}{\ensuremath{{\rm d}}}

\newcommand{\del}{\ensuremath{\partial}}
\newcommand{\Del}{\ensuremath{\nabla}}

\newcommand{\ba}{\begin{eqnarray}}
\newcommand{\ea}{\end{eqnarray}}
\begin{document}

\bigskip
\begin{titlepage}

\begin{flushright}
UUITP-06/04\\
hep-th/0402192
\end{flushright}

\vspace{1cm}

\begin{center}
{\Large\bf  Type 0A 2D Black Hole Thermodynamics and the Deformed Matrix Model\\}

\end{center}
\vspace{3mm}

\begin{center}

{\large
Ulf   H.\   Danielsson{$^1$},   James   P.  Gregory{$^2$},  Martin  E.
Olsson{$^3$}, \\ Peter Rajan{$^4$} and Marcel Vonk{$^5$}} \\
\vspace{5mm}

Institutionen f\"or Teoretisk Fysik, Box 803, SE-751 08
Uppsala, Sweden

\vspace{5mm}

{\tt
{$^1$}ulf.danielsson@teorfys.uu.se \\
{$^2$}james.gregory@teorfys.uu.se \\
{$^3$}martin.olsson@teorfys.uu.se \\
{$^4$}peter.rajan@teorfys.uu.se \\
{$^5$}marcel.vonk@teorfys.uu.se \\
}

\end{center}
\vspace{5mm}

\begin{center}
{\large \bf Abstract}
\end{center}
\noindent
Recently,  it  has  been  proposed  that  the  deformed  matrix  model
describes  a  two-dimensional  type  0A  extremal  black hole. In this
paper,  the  thermodynamics  of 0A charged non-extremal black holes is
investigated.  We  observe that the free energy of the deformed matrix
model  to  leading  order  in  $1/q$ can be seen to agree to that of the
extremal  black  hole.  We  also  speculate on how the deformed matrix
model  is  able  to  describe the thermodynamics of non-extremal black
holes.

\vfill
\begin{flushleft}
February 2004
\end{flushleft}
\end{titlepage}
\newpage

\section{Introduction}

Recently,   the   work  in  \cite{McGreevy:2003st,  Takayanagi:2003mm,
Douglas:2003nh}  has revived the interest in matrix model descriptions
of  noncritical  string  theories in two dimensions. In particular, it
has  been  argued in \cite{Takayanagi:2003mm, Douglas:2003nh} that the
two-dimensional  type  0A and type 0B string theories each are dual to
matrix models which are nonperturbatively stable.

The  so-called  deformed  matrix model which was suggested to describe
the  two-dimensional  type  0A  string  theory  had  appeared  in  the
literature  long  before these developments -- in \cite{Jevicki:1994dm}
it  was  proposed  that  this  matrix  model describes two-dimensional
string  theory in a black hole background. For further developments of
this see [5--10]. Very  recently
\cite{Danielsson:2003mm,  Gukov:2003fb},  it  was  argued that both of
these identifications are essentially correct, and that type 0A string
theory  in  the  presence  of  a  background flux $q$ can be viewed as
describing an extremal two-dimensional black hole of charge $q$.

In  this paper, we give further evidence for this claim by calculating
the thermodynamical properties of non-extremal two-dimensional charged
black  holes. In particular, we calculate the free energy, entropy and
temperature  of these black holes. We show that in the extremal limit,
this  free  energy  exactly matches the matrix model free energy. This
supports  the  observation  made  in  \cite{Gukov:2003fb} where it was
found  that  the  ADM  mass of the extremal black hole equals the free
energy of the 0A matrix model.

The  next  natural  question  to ask is whether there is also a matrix
model description of non-extremal two-dimensional black holes. In this
respect, the cutoff dependence we find on the space-time side is quite
suggestive.  Motivated  by this, we discuss a possible modification of
the deformed matrix model based on the work in~\cite{Kazakov:2001mm}.

This  paper  is organized as follows. In section \ref{sec:space-time},
we  discuss  the  thermodynamics  of the two-dimensional charged black
hole and calculate the free energy by evaluating the Euclidean action.
In  section  \ref{sec:matrixmodel},  we  begin by calculating the free
energy  of  the  type  0A matrix model, finding a result which exactly
matches  the  result  of  the  previous  section in the extremal case.
Subsequently,  we  discuss in more detail the relation between the two
sides  of  the  duality.  In  particular, we describe how the deformed
matrix  model  could  be  modified  in  order  to take the temperature
dependence of the non-extremal black hole into account. We conclude in
section \ref{sec:conclusions}. \\

\noindent
{\bf  Note added.} In preparing this paper for submission, we received
the  paper  \cite{Davis:2004bh}, where the thermodynamic properties of
the  two-dimensional charged black hole are also calculated. While our
results  on the thermodynamics on the space-time side agree with those
in  \cite{Davis:2004bh},  the  conclusions  on  the comparision to the
matrix model seem to differ.

\sect{Free Energy and Thermodynamics of the Black Hole} \label{sec:space-time}

\subsection{The Black Hole Solution}

The starting point for our analysis will be the space-time effective action for
type 0A string theory \cite{McGuigan:1992cb, Berkovits:2001sb,
  Danielsson:2003mm, Gukov:2003fb} in the presence of a RR flux $q$
and zero tachyon field,\footnote{The sign of the action varies in the
  literature but this choice of sign depends on the definition of the
  spatial coordinate in the measure.}
\be \label{eq:stringact}
 S = - \int \d^2 x \sqrt{-g} \left[ e^{-2\Phi} \left( \mathcal{R} + 4 (\nabla
   \Phi)^2 + 4k^2 \right) + \La \right],
\ee
where $\Phi$ is the dilaton field and the constants $k$ and $\Lambda$
are given by\footnote{We note that, in comparison to the actions given
in~\cite{Gukov:2003fb,Davis:2004bh}, our value of $\Lambda$ differs by a factor of 2.
This is because we choose only to include contributions to the $\Lambda$ term from
either the electric or magnetic sector of D0 branes, which allows for a generalization
to a non-zero tachyon field.  For a discussion of this point we refer the reader to,
for example,~\cite{Douglas:2003nh}.}
\be \label{eq:bhparameters}
k = \sqrt{\frac{2}{\al'}} \qquad\qquad \La = \frac{-q^2}{4\pi\al'}.
\ee
The equations of motion following from this action are
\ba
\mathcal{R}_{\mu\nu}-\frac{\La}{2}e^{2\Phi}g_{\mu\nu}+2\Del_\mu\Del_\nu\Phi
& = & 0, \\
\mathcal{R} + 4k^2+4\Del^2\Phi-4(\Del\Phi)^2 & = & 0.
\ea
Denoting the space-time coordinates by $(t,\phi)$, one finds the
following one-parameter family of solutions~\cite{Berkovits:2001sb,
  Danielsson:2003mm, Gukov:2003fb}
\ba
 ds^2 & = & - l(\phi) \d t^2 + \frac{1}{l(\phi)} \d \phi^2 \\
\Phi & = & - k \phi,
\ea
where
\be
 l(\phi) = 1 - e^{-2k\phi}\left(\frac{m}{k} - \frac{\La}{2k}\phi \right).
\ee
The nature of such solutions depends on the value of the parameter
$m$ in relation to its critical value
\be
m_\textrm{ext} = \frac{\La}{4k}\ln\left(\frac{-\La}{4k^2e}\right).
\ee
For $m < m_\textrm{ext}$ the space-time has a naked singularity and these
solutions will not be of interest to us.  For $m \ge m_\textrm{ext}$ the
solution represents a charged black hole whose horizon, $\phi_H$, is
given by the largest zero of $l(\phi)$.  Whilst $l(\phi)$ generically
has two zeros for the black hole space-time, these coincide when the
lower bound on $m$ is saturated and at this point the black hole is
extremal.

\subsection{Basic Black Hole Thermodynamics}

Our two-dimensional black holes are parametrized by the mass parameter $m$.
However, it often becomes easier to parametrize by the horizon radius
$\phi_H$ instead. Using $l(\phi_H)=0$, these two quantities are
related by the one-to-one correspondence
\be
m = k e^{2k\phi_H} +\frac{\La}{2}\phi_H.
\ee
The temperature of the black hole can be defined by moving to the
Euclidean section $t\to i\tau$.  In so doing the horizon develops a
bolt singularity which must subsequently be removed by compactifying
the Euclidean time direction $\tau$ according to~\cite{Gibbons:1977ai}
\be
\tau\sim\tau+\beta \quad \textrm{where} \quad \beta =
\frac{4\pi}{l'(\phi)}.
\ee
The temperature of the black hole is then defined by
\be \label{eq:bhtemp}
T = \frac{1}{\beta} = \frac{1}{8\pi k}\left( 4k^2 +\La
e^{-2k\phi_H}\right)
\ee
The relation now of course allows us to parametrize the black hole
space-time in terms of $T$ instead by using the one-to-one mapping
\be
\phi_H = -\frac{1}{2k}\ln\left(\frac{8\pi k T -4k^2}{\La}\right).
\ee
In what follows we will refer to horizon radius of the extremal black
hole $\phi_H(T=0)$ as $\phi_\textrm{ext}$.

The mass of the black hole space-time can be evaluated via the
ADM formalism (see for example~\cite{Liebl:1996hr}).  This requires a
reference space-time for which we take the extremal black hole. The ADM
energy is then given by
\ba \label{eq:ADM}
E_\textrm{ADM} = M - M_\textrm{ext} & = &
\left[-4ke^{2k\phi}\left(l(\phi)-\sqrt{l(\phi)l_\textrm{ext}(\phi)}\right)
\right]_{\phi\to\infty} \nonumber \\
& = & \left(2m-\Lambda \phi_c\right) -
\left(2m_\textrm{ext}-\Lambda \phi_c\right),
\ea
where $M = 2m - \Lambda \phi_c$ is the mass of the black hole space-time
including a divergent volume term.  We have regulated this
contribution with an infrared cutoff $\phi_c$ in the spatial direction
$\phi$ and ignored terms which vanish in the large $\phi_c$ limit.

To completely determine the thermodynamics of the black hole it now
remains to calculate the free energy from which one can also derive
the entropy.

\subsection{Free Energy from the Euclidean Action}

Our objective now is to calculate the free energy by using Euclidean
Quantum Gravity techniques.  Of course, one may be tempted to
calculate the entropy and free energy directly by integrating up the
second law of thermodynamics, but the integration constants which
arise could be of importance so we will avoid this problem in what
follows.

The free energy, $F$, can be defined in terms of the on-shell value of
the Euclidean action, $I$, through the simple relation
\be
F= \frac{I}{\beta}.
\ee
We thus proceed to evaluate the Euclidean action and the remaining
thermodynamic properties of our solution will be revealed.

In moving to the Euclidean section, $t \to i\tau$, one must supplement
an action with a Gibbons-Hawking boundary term which allows the use of
a well-defined variational principle~\cite{Gibbons:1977ai}.  In doing
so our complete Euclidean action now becomes
\be
I = -\int_\mathcal{M} \sqrt{g}\left[e^{-2\Phi} \left( \mathcal{R} + 4
  (\Del \Phi)^2 + 4k^2 \right) + \La\right] -
2\int_{\del\mathcal{M}}\sqrt{h}\,e^{-2\Phi} K,
\ee
where $h_{a b}$ is the induced metric on the boundary and $K$ is the
trace of the second fundamental form.

Formally  evaluating the action for our solution will give infinity as
we  intend to evaluate over all Euclidean space, {\it i.e.} the entire
region  outside  of  the  black  hole  horizon.  To  deal  with this a
background subtraction is required. In choosing a reference background
it  must  be one which belongs to the ensemble of black hole solutions
that  are  under  consideration,  {\it i.e.} in this case it must also
have  charge  $q$. Therefore the natural background to consider is the
extremal  black  hole  defined  by $m=m_\textrm{ext}$. Our regularized
action then becomes
\ba
I & = & -\left(\int_{\mathcal{M}}-\int_{\mathcal{M}_\textrm{ext}}\right)
\sqrt{g}\left[e^{-2\Phi} \left( \mathcal{R} + 4 (\nabla \Phi)^2 + 4k^2
  \right) + \La\right] \nonumber \\
&& \quad - 2
\left(\int_{\del\mathcal{M}}-\int_{\del\mathcal{M}_\textrm{ext}}\right)
\sqrt{h}\,e^{-2\Phi} K.
\ea
In order to match the black hole with the extremal black hole
reference background at spatial infinity the limits of integration
must be chosen with care.  Our intention is to evaluate over all
space-time, but this is achieved by formally integrating up to some
cutoff $R$ in the $\phi$ coordinates for both space-times and then
letting $R\to\infty$.  The $\tau$ integration then proceeds as
follows.  Whilst for the black hole we integrate $\tau$ from $0$ to
$\beta$, the geometry of the reference background must match the black
hole geometry at the cutoff surface $\phi=R$.  This requires us to
choose a period of integration for the $\tau$ coordinate of the
extremal black hole given by the following relation\footnote{This is
  valid since for the extremal black hole there is no bolt singularity
  at the horizon and so one may formally assign any desired period to the
  Euclidean time~\cite{Gibbons:1995te, Hawking:1995ea}.}
\be
\beta\sqrt{g^{BH}_{\tau\tau}(R)} =
\beta_\textrm{ext}\sqrt{g^\textrm{ext}_{\tau\tau}(R)}.
\ee
Evaluating $\beta_\textrm{ext}$ at large $R$ we find
\be
\frac{\beta_\textrm{ext}}{\beta} = 1 - \frac{m-m_\textrm{ext}}{2k}e^{-2kR} +
\mathcal{O}\left(e^{-4kR}\right).
\ee

Given that the limits of integration are now established we may turn
to the integrands.  Evaluating the bulk integrand we find
\be
\mathcal{L}_\textrm{bulk} = 3\La + 8 k^2e^{2k\phi}.
\ee
The boundary integrand needs a little bit of technology.  We pick a
unit normal to the boundary $n^\mu = (0, 1/\sqrt{g_{\phi\phi}})$.
Then $K_{ab}$ is the Lie derivative of the induced metric with respect
to this normal, {\it i.e.}
\be
K_{ab}=h^c_{(a}h^d_{b)} \del_c n_d.
\ee
In our case the only nonzero component of $K$ is the
$\tau\tau$-component and the trace of $K$ can be subsequently found,
\be
K_{\tau\tau}=\frac{1}{2}\sqrt{l(\phi)}l'(\phi) \quad \Rightarrow \quad
\sqrt{h}\,K=\frac{l'(\phi)}{2}.
\ee
We can now evaluate the integrand of the boundary integral to obtain
\be
\mathcal{L}_\textrm{bdry} = \bar{m}-\frac{R\La}{2}+\frac{\La}{2k},
\ee
where $\bar{m}$ is $m$ or $m_\textrm{ext}$ respectively.

We now have all the pieces we need to evaluate the Euclidean action of
our solution
\ba
I & = & -\left(\int_0^\beta\d\tau \int_{\phi_H}^{R}\d\phi -
\int_0^{\beta_\textrm{ext}}\d\tau
\int_{\phi_\textrm{ext}}^{R}\d\phi\right)\mathcal{L}_\textrm{bulk} -
2\left(\int_0^\beta\d\tau - \int_0^{\beta_\textrm{ext}}\d\tau\right)
\mathcal{L}_\textrm{bdry} \nonumber \\
& = & -\beta\left[ \left(3\La R+4k e^{2kR}\right)
  \left(1-\frac{\beta_\textrm{ext}}{\beta}\right) - \left(3\La \phi_H + 4k
  e^{2k\phi_H}\right) +\frac{\beta_\textrm{ext}}{\beta}\left(3\La
  \phi_\textrm{ext} + 4k
  e^{2k\phi_\textrm{ext}}\right)\right] \nonumber \\
&& \quad - 2\beta\left[m-\frac{R\La}{2}+\frac{\La}{2k} -
  \frac{\beta_\textrm{ext}}{\beta}
  \left(m_\textrm{ext}-\frac{R\La}{2}+\frac{\La}{2k}\right)\right].
\ea
Taking $R\to\infty$ we obtain the action of our black hole space-time
relative to the extremal black hole reference background
\be
I = \beta \La \left(\phi_H-\phi_\textrm{ext}\right).
\ee
The free energy of the solution now follows.  Given that we are
considering an ensemble of fixed charge black hole solutions, what we
have actually calculated is the Helmholtz free energy for the
canonical ensemble.  We therefore know that this is given by the
following relation
\be \label{eq:free}
F = \frac{I}{\beta} = E - TS
\ee
where $E = M - M_\textrm{ext}$.

\subsection{Black Hole Entropy}

In order to calculate the entropy in the simplest manner the free
energy should first be expressed in terms of temperature alone
\be
F = -\frac{\La}{2k}\ln\left(1-\frac{2\pi}{k}T\right).
\ee
Notice that the free energy for the extremal black hole is zero as we
would expect since we have the extremal black hole as a reference
background.

The black hole entropy then follows from the standard thermodynamic
relation relating Helmholtz free energy to entropy
\ba \label{eq:entropy}
S = -\frac{\del F}{\del T} & = &
-\frac{\pi\La}{k^2}\left(1-\frac{2\pi}{k}T\right)^{-1} \nonumber \\
& = & 4\pi e^{2k\phi_H}.
\ea
It is important to observe that the entropy of the extremal black hole
is non-zero.  Furthermore, we comment that this entropy can indeed be
obtained by integrating up the second law of thermodynamics given our
value for the temperature and the ADM mass of the black hole.  Expressing
both as functions of $\phi_H$ and integrating over what seems the most
natural range for this radial coordinate, $[\phi_H,\infty]$, the
correct constant of integration is obtained.

As an aside, note that we can express the entropy (\ref{eq:entropy})
in a familiar form,
\be
S=4\pi e^{2k\phi_{H}} = \frac{4\pi}{e^{2\Phi_{H}}}=
\frac{A}{4 G_{N}},
\ee
where Newton's constant, $G_{N}$, is expressed in terms of the dilaton
at the horizon, $G_{N}=e^{2\Phi_{H}}/16\pi$.  Here we are using the
natural value $A=1$ for the area of the black hole horizon as seen
from one side of the singularity.

\subsection{Non-zero Free Energy of the Extremal Black Hole}

Given the fact that the free energy we have calculated is the
Helmholtz free energy, we have been able to express this in terms of
the basic thermodynamic quantities of the black hole
space-time~(\ref{eq:free}):
\be
F = M - M_\textrm{ext} - TS.
\ee
Obviously such a free energy includes a contribution from the
reference space-time of our subtraction scheme, but this subtraction
may be easily removed.  We can calculate the free energy of the black
hole space-time alone by merely adding back on the contribution from
the mass of the extremal black hole
\be
F_\textrm{BH} = F + M_\textrm{ext} \equiv M - TS.
\ee
The resulting free energy is given by
\ba \label{eq:bhfreeenergy}
F_\textrm{BH} = \La\phi_H -\La\phi_\textrm{ext} +2m_\textrm{ext} -
\Lambda \phi_c & = & \La\left(\phi_H - \phi_c\right) - \frac{\La}{2k}
\nonumber \\
& = & -\frac{\Lambda}{2k}\ln\left( e^{1+2\left(\Phi_H-\Phi_c\right)}
\right),
\ea
where in the last line we have used $\Phi=-k\phi$.  This free
energy can also be expressed in terms of the temperature as
\be
F_{\textrm{BH}} = -\frac{\La}{2k}\ln\left(\frac{-4k^2e}{\La}\left[
  1-\frac{2\pi}{k}T \right]e^{-2\Phi_c}\right).
\ee
This is obviously the free energy presented earlier up to the additive
constant of the extremal black hole mass so the entropy is unaffected.

Most importantly to us now, we have a non-zero free energy for the
case of the extremal black hole which we shall demonstrate is
reproduced in the matrix model,
\be \label{eq:extremalfree}
F_{\textrm{BH}}(T=0) =
-\frac{\La}{2k}\ln\left(\frac{-4k^2e}{\La}\,e^{-2\Phi_c} \right)=
-\frac{q^2}{16\pi}\sqrt{\frac{2}{\al'}} \ln\left(\frac{q^2}{32\pi
  e}\,e^{2\Phi_c}\right),
\ee
where the definitions (\ref{eq:bhparameters}) have been inserted.

\sect{The Matrix Model Free Energy} \label{sec:matrixmodel}

Let  us  now  turn to the deformed matrix model and calculate the free
energy  there.\footnote{Such calculations have previously been done in,
{\it e.g.}, \cite{Demeterfi:1993sq}.} We
will  find  that,  to  leading order in $1/q$, the extremal value of the
black  hole  free  energy  (\ref{eq:extremalfree}) precisely matches the
matrix model free energy calculated below.

We  shall  also  discuss  how  the  matrix  model  can  be modified in order to
reproduce  the $T$-dependent free energy representing the non-extremal
black hole.

\subsection{Free Energy from the Matrix Model}

It is well known that the free energy of the matrix model can be written as
\begin{equation}
F_{\textrm{MM}}=\int_{-\vep_{c} }^{-\mu }\varepsilon \rho \left( \varepsilon \right)
d\varepsilon ,
\end{equation}
where the density of the number of states is given by
\begin{equation}
\rho \left( \varepsilon \right) = - \frac{1}{\pi }\mathrm{Im}\sum_{n=0}\frac{1}{E_
{n}+\varepsilon }
\end{equation}
at zero temperature. $-\mu $ is the position of the Fermi sea, and $-\vep_{c}
$ is a cutoff.

Following \cite{Danielsson:1993mm} we may formally continue $\alpha ^{\prime
}\rightarrow -\alpha ^{\prime }$ in order to have a right side up harmonic
oscillator. We then have the following Schr\"{o}dinger equation to solve:
\begin{equation}
\left( -\frac{1}{2}\frac{d^{2}}{dx^{2}}+\frac{1}{4\alpha ^{\prime }}
x ^{2}+\frac{\eta }{2x^{2}}\right) \psi \left( x\right) =E\psi
\left( x\right) ,
\end{equation}
where $\eta =q^{2}-1/4$. Using this, the energy eigenvalues of the original problem can be shown to be
\begin{equation}
E_{n}=\frac{i}{2\sqrt{2\alpha ^{\prime }}}\left( 2n+1+2a_{n}\right) ,
\end{equation}
where
\begin{equation}
a_{n}=-(-1)^{n}\left( -\frac{1}{2}+\sqrt{\frac{1}{4}+\eta }\right) .
\end{equation}
For $q\geq 0$, we must throw away the even parity solutions leaving only the
odd levels given by
\begin{equation}
E_{2l+1}=\frac{i}{2\sqrt{2\alpha ^{\prime }}}\left( 4l+3-1+2q\right) =\frac{i}
{\sqrt{2\alpha ^{\prime }}}\left( 2l+1+q\right) .
\end{equation}
To leading order in large $q$ we find
\begin{equation}
\rho \left( \varepsilon \right) =-\frac{1}{\pi }\mathrm{Im}\sum_{l=0}\frac{1}
{\frac{i}{\sqrt{2\alpha ^{\prime }}}
\left( 2l+1+q\right) +\varepsilon }\sim
-\frac{1}{2\pi }\sqrt{\frac{\alpha ^{\prime }}{2}}\ln \left( \frac{q^{2}}{4}+
\frac{\alpha ^{\prime }}{2}\varepsilon ^{2}\right) +\mathrm{const.,}
\end{equation}
where the constant depends on the cutoff $\vep_{c} $ but is independent
of $q$.
Using this we find a free energy (at $\mu=0$) given by
\begin{equation} \label{eq:matrixfree}
F_{\textrm{MM}}=-\frac{q^{2}}{16\pi }\sqrt{\frac{2}{\alpha ^{\prime }}}\ln \left(
  \frac{q^{2}}{4e}
\frac{2}{\alpha ^{\prime }\vep_{c} ^{2}}\right) +...,
\end{equation}
up to terms of higher order in $1/\vep_{c} $ or independent of $q$.

\subsection{Comparing the Deformed Matrix Model with the Black Hole} \label
{sec:nonextremal}

If we compare (\ref{eq:matrixfree}) with the free
energy in equation (\ref{eq:extremalfree}), derived from the
space-time point of view, we  see that the numerical coefficient in front of
the logarithm is in
perfect agreement. That is, the cutoff independent non-analytical (in
$q$) parts of the free
energies are identical. The agreement carries over also to the analytic cutoff
dependent part of the expression provided that we relate the spatial
space-time cutoff $\phi_{c}$ to the energy cutoff $\vep_{c}$ in the matrix model.
The latter can also be understood as an IR-cutoff in the matrix eigenvalue
coordinate $x$. Even though the cutoffs are related in a natural
way, it is well known that the relation between the matrix eigenvalue
coordinate and space-time is quite complicated and given by a
nonlocal integral transform. Nevertheless, it is reassuring to see
the correspondence.

While  it  is gratifying that we obtain this agreement on the level of
the  free  energy  at zero temperature, it would be nice to be able to
reproduce  the  full  temperature  dependent  free  energy  using  the
deformed  matrix  model.  In  this  context  one  should note that the
temperature  dependence  of (\ref{eq:bhfreeenergy}) is of an extremely
simple  form. Indeed, all of the temperature dependence comes from the
way  in  which  the  spatial  volume  depends  on  the position of the
horizon.\footnote{One  is  tempted  to  write the cutoff on the matrix
model  side  as $x_{H}^{4}/x_{c}^{4}$ in a similar fashion.} This can,
possibly,   be   a  clue  towards  a  complete  understanding  of  the
correspondence from the point of view of the matrix model.

Unfortunately,  it  is  not  enough  to  turn  on a temperature in the
deformed  matrix  model  simply  by  compactifying Euclidean time. The
resulting  free energy will not be of the form (\ref{eq:bhfreeenergy})
and  will  not  give  rise  to  the  correct entropy. The most natural
interpretation of this observation is that the naive way of turning on
a  temperature  just corresponds to placing the extremal black hole in
an external heat bath without changing the mass.

Instead it is reasonable to expect that we need yet another parameter to
deform the matrix model. An interesting candidate is found if we make use of
the non-singlet states of the matrix model as is done in
\cite{Kazakov:2001mm}. It is claimed in that paper that the fugacity of the
vortices
is a measure of the mass of a Witten type uncharged black hole \cite{Witten:1991st}. In our
case we expect the fugacity to be a measure of the departure from
extremality, and, as a consequence, the
calculations of \cite{Kazakov:2001mm} would have
to be redone for the deformed matrix model. In principle we would then
have two free parameters: the fugacity and the temperature. In order
to reproduce the correct thermodynamics, however, there needs to be an
extra relation analogous to the space-time relation between the temperature and the mass.
It would be interesting to see how
this comes about in the matrix model.\footnote{One should note that in case
of the
uncharged black hole the temperature is independent of the
mass. Compare (\ref{eq:bhtemp}) which for $q=0$ gives $T=\frac{k}{2\pi}$.}

\sect{Conclusions} \label{sec:conclusions}

In   this   paper   we   have  discussed  the  thermodynamics  of  the
two-dimensional  type  0A  black  hole.  In  particular,  using  the
Euclidean  approach,  we  calculated the (non-extremal) classical free
energy (\ref{eq:bhfreeenergy}), from which all the other quantities of
interest follow in the usual way.

In  comparing  the  space-time  results  to the conjectured dual $0$A
matrix  model  we  found  that the free energies agree in the extremal
limit.  It  would  quite  naturally  be very interesting to see if the
agreement between the free energies still holds away from extremality.
In  particular  one  would like to be able to calculate the black hole
entropy  directly  from  the  matrix  model.  The correct matrix model
description of this non-extremal black hole is still an open question.

However,  we  believe  that the evidence presented here (see also
\cite{Danielsson:2003mm,Gukov:2003fb}) in   favour  of  the  duality
between  the deformed matrix model and the $0$A black hole is
an important step towards a better understanding of this issue.

\section*{Acknowledgments}
We  would  like  to thank A. Padilla for correspondence. UD is a Royal
Swedish  Academy of Sciences Research Fellow supported by a grant from
the  Knut and Alice Wallenberg Foundation. The work was also supported
by the Swedish Research Council (VR).


\begin{thebibliography}{99}

\bibitem{McGreevy:2003st}
J.~McGreevy and H.~Verlinde,
{\em Strings from tachyons: the $c=1$ matrix reloaded,}
\jhep{0312}{2003}{054} [hep-th/0304224].

\bibitem{Takayanagi:2003mm}
T.~Takayanagi and N.~Toumbas,
{\em A matrix model dual of type 0B string theory in two dimensions,}
\jhep{0307}{2003}{064} [hep-th/0307083].

\bibitem{Douglas:2003nh}
M.~R.~Douglas, I.~R.~Klebanov, D.~Kutasov, J.~Maldacena, E.~Martinec and
N.~Seiberg,
{\em A new hat for the $c=1$ matrix model}
[hep-th/0307195].

\bibitem{Jevicki:1994dm}
A.~Jevicki and T.~Yoneya,
{\em A deformed matrix model and the black hole background in two-dimensional
string theory,}
\npb{411}{1994}{64-96} [hep-th/9305109].


\bibitem{Danielsson:1993mm}  U.~H.~Danielsson, {\it A matrix model black
hole,} \npb{410}{1993}{395-406} [hep-th/9306063].

\bibitem{Demeterfi:1993sq}
K.~Demeterfi and J.~P.~Rodrigues,
{\em States and quantum effects in the collective field theory of a deformed
matrix model,}
\npb{415}{1994}{3-28} [hep-th/9306141].

\bibitem{Demeterfi:1993es}  K.~Demeterfi, I.~R.~Klebanov and
J.~P.~Rodrigues, {\it The exact S matrix of the deformed c = 1 matrix model,}
\prl{71}{1993}{3409-3412} [hep-th/9308036].

\bibitem{Danielsson:1993dm}  U.~H.~Danielsson, {\it The deformed matrix model
at finite radius and a new duality symmetry,} \plb{325}{1994}{33-38} [hep-th/9309157].

\bibitem{Danielsson:1994ts}  U.~H.~Danielsson, {\it Two-dimensional string
theory, topological field theories and the deformed matrix model,}
\npb{425}{1994}{261-276} [hep-th/9401135].

\bibitem{Danielsson:1994ss}  U.~H.~Danielsson, {\it The scattering of strings
in a black hole background,} \plb{338}{1994}{158-164}
[hep-th/9405052].


\bibitem{Danielsson:2003mm}
U.~H.~Danielsson,
{\em A matrix model black hole: act II,}
\jhep{0402}{2004}{067} [hep-th/0312203].

\bibitem{Gukov:2003fb}
S.~Gukov, T.~Takayanagi and N.~Toumbas,
{\em Flux backgrounds in 2D string theory,}
\jhep{0403}{2004}{017} [hep-th/0312208].

\bibitem{Kazakov:2001mm}
V.~Kazakov, I.~K.~Kostov and D.~Kutasov,
{\em A matrix model for the two dimensional black hole,}
\npb{622}{2002}{141-188} [hep-th/0101011].

\bibitem{Davis:2004bh}
J.~Davis, L.~A.~Pando Zayas and D.~Vaman,
{\em On black hole thermodynamics of 2-D type 0A,}
\jhep{0403}{2004}{007} [hep-th/0402152].

\bibitem{McGuigan:1992cb}
M.~D.~McGuigan, C.~R.~Nappi and S.~A.~Yost,
{\em Charged black holes in two-dimensional string theory,}
\npb{375}{1992}{421-452} [hep-th/9111038].

\bibitem{Berkovits:2001sb}
N.~Berkovits, S.~Gukov and B.~C.~Vallilo,
{\em Superstrings in 2D backgrounds with R-R flux and new extremal black
holes,}
\npb{614}{2001}{195-232} [hep-th/0107140].

\bibitem{Gibbons:1977ai}
G.~Gibbons, S.~W.~Hawking,
{\em Action integrals and partition functions in quantum gravity,}
\prd{15}{1977}{2752-2756}.

\bibitem{Liebl:1996hr}
H.~Liebl, D.~V.~Vassilevich and S.~Alexandrov,
{\em Hawking radiation and masses in generalized dilaton theories,}
\cqg{14}{1997}{889-904} [gr-qc/9605044].

\bibitem{Gibbons:1995te}
G.~Gibbons, R.~Kallosh,
{\em Topology, entropy and Witten index of dilaton black holes,}
\prd{51}{1995}{2839-2862} [hep-th/9407118].

\bibitem{Hawking:1995ea}
S.~W.~Hawking, G.~T.~Horowitz and S.~F.~Ross,
{\em Entropy, area, and black hole pairs,}
\prd{51}{1995}{4302-4314} [gr-qc/9409013].


\bibitem{Witten:1991st}  E.~Witten, {\it On string theory and black holes,}
\prd{44}{1991}{314-324}.


\end{thebibliography}
\end{document}